\documentclass[12pt]{article}
\usepackage{amsmath}
\usepackage{amssymb}

\input{tcilatex}
\begin{document}

\begin{center}
\textbf{COMPLEXIFIED SIGMA MODEL AND DUALITY}

\smallskip \ 

\smallskip \ 

\smallskip \ 

J. A. Nieto \footnote{%
nieto@uas.uasnet.mx, janieto1@asu.edu}

\smallskip \ 

\textit{Facultad de Ciencias F\'{\i}sico-Matem\'{a}ticas de la Universidad
Aut\'{o}noma} \textit{de Sinaloa, 80010, Culiac\'{a}n Sinaloa, M\'{e}xico.}

\smallskip

and

\smallskip

\textit{Mathematical, Computational \& Modeling Sciences Center, Arizona
State University, PO Box 871904, Tempe AZ 85287, USA}

\bigskip \ 

\bigskip \ 

\textbf{Abstract}

\smallskip
\end{center}

We show that the equations of motion associated with a complexified
sigma-model action do not admit manifest dual SO(n,n) symmetry. In the
process we discover new type of numbers which we called `complexoids' in
order to emphasize their close relation with both complex numbers and
matroids. It turns out that the complexoids allow to consider the analogue
of the complexified sigma-model action but with (1+1)-worldsheet metric,
instead of Euclidean-worldsheet metric. Our observations can be useful for
further developments of complexified quantum mechanics.

\bigskip \ 

\bigskip \ 

\bigskip \ 

\bigskip \ 

\bigskip \ 


Pacs numbers: 04.60.-m, 04.65.+e, 11.15.-q, 11.30.Ly

May, 2011

\newpage

Recently, Witten [1] has introduced new ideas on quantum mechanics via a
complaxification of the Feynman path integral of ordinary quantum mechanics.
It turns out that one of the key formulas in these new developments is the
generalized Cauchy-Riemann conditions

\begin{equation}
\frac{\partial x^{\mu }}{\partial \sigma }=-g^{\mu \nu }\frac{\partial
x^{\alpha }}{\partial \tau }b_{\nu \alpha },  \tag{1}
\end{equation}%
which can be obtained starting with the Morse theory flow equation (see Ref.
[1] for details). (By convenience, in writing (1) we have modified the
notation of Ref. [1].) The metric $g_{\mu \nu }$ and the nonsymmetric form $%
b_{\nu \alpha }=-b_{\nu \alpha }$in (1) are choosing by requiring that the
quantity $b_{~\alpha }^{\mu }=g^{\mu \nu }b_{\nu \alpha }$ satisfies 
\begin{equation}
b_{~\beta }^{\mu }b_{~\tau }^{\beta }=-\delta _{\tau }^{\mu }.  \tag{2}
\end{equation}%
Under this condition $b_{~\alpha }^{\mu }$ turns out to be a direct sum of $%
2\times 2$-block matrices of the form

\begin{equation}
\varepsilon _{ij}\equiv \left( 
\begin{array}{cc}
0 & 1 \\ 
-1 & 0%
\end{array}%
\right) .  \tag{3}
\end{equation}%
In each block, the equation (1) gives

\begin{equation}
\begin{array}{c}
\frac{\partial u}{\partial \sigma }=\frac{\partial v}{\partial \tau }, \\ 
\\ 
\frac{\partial v}{\partial \sigma }=-\frac{\partial u}{\partial \tau },%
\end{array}
\tag{4}
\end{equation}%
which are the familiar Cauchy-Riemann equations in two dimensions . This
means that the quantity $f=u+iv$ is a holomorphic function of $z=\sigma
+i\tau $. (Here, we are assuming that the partial derivatives in (1) are
continuous.)

Consider the first order differential $\omega =\omega _{i}d\xi ^{i}=\omega
_{1}d\sigma +\omega _{2}d\tau $, where $\omega _{1}$ and $\omega _{2}$ are
complex valued functions. If one introduces the conjugate differential $\ast
\omega =\ast \omega _{i}d\xi ^{i}=-\omega _{2}d\sigma +\omega _{1}d\tau $
one can verify that%
\begin{equation}
\ast \omega _{i}=-\varepsilon _{ij}\omega ^{j},  \tag{5}
\end{equation}%
where $\omega ^{i}=\delta ^{ij}\omega _{j}$, with

\begin{equation}
\delta _{ij}\equiv \left( 
\begin{array}{cc}
1 & 0 \\ 
0 & 1%
\end{array}%
\right) .  \tag{6}
\end{equation}%
Observe that $\ast ^{2}=-1$. It can be shown that $\omega $ is holomorphic
differential on a specified region if and only if $d\omega =0$ and $\ast
\omega =-i\omega $. Locally this is equivalent to 
\begin{equation}
\ast df=-idf,  \tag{7}
\end{equation}%
for some function $f=u+iv$. It turns out that (7) is equivalent to

\begin{equation}
\ast dv=du.  \tag{8}
\end{equation}%
This means that $du$ and $dv$ are conjugate differentials. In fact, one can
show that the condition (8) is equivalent to the Cauchy-Riemann conditions
(4) (see Ref. [2] for details). These basic observations allow us to write
(1) in the form

\begin{equation}
\mathcal{U}=dx^{\mu }-\ast b_{~\nu }^{\mu }dx^{\nu }=0.  \tag{9}
\end{equation}%
Considering (5) one sees that this expression implies

\begin{equation}
\mathcal{U}^{i\mu }=\delta ^{ij}\frac{\partial x^{\mu }}{\partial \xi ^{j}}%
+g^{\mu \alpha }b_{\alpha \nu }\varepsilon ^{ij}\frac{\partial x^{\nu }}{%
\partial \xi ^{j}}=0,  \tag{10}
\end{equation}%
which can also be written as

\begin{equation}
\mathcal{U}_{\mu }^{i}=\delta ^{ij}\frac{\partial x^{\nu }}{\partial \xi ^{j}%
}g_{\mu \nu }+\varepsilon ^{ij}\frac{\partial x^{\nu }}{\partial \xi ^{j}}%
b_{\mu \nu }=0.  \tag{11}
\end{equation}%
It is interesting to write $\mathcal{U}_{\mu }^{i}$ in the form

\begin{equation}
\mathcal{U}_{\mu }^{i}=\mathcal{G}_{\mu \nu }^{ij}\frac{\partial x^{\nu }}{%
\partial \xi ^{j}},  \tag{12}
\end{equation}%
with

\begin{equation}
\mathcal{G}_{\mu \nu }^{ij}=\delta ^{ij}g_{\mu \nu }+\varepsilon ^{ij}b_{\mu
\nu }.  \tag{13}
\end{equation}%
We recognize in (13) the generalized metric proposed in [3].

Consider now the bosonic action [1]

\begin{equation}
S=\frac{1}{2}\int_{M^{2}}\mathcal{F}^{\mu }\wedge \mathcal{U}_{\mu }+\frac{1%
}{4}\int_{M^{2}}\mathcal{F}^{\mu }\wedge \ast \mathcal{F}^{\nu }g_{\mu \nu }.
\tag{14}
\end{equation}%
Here, $\mathcal{F}^{\mu }$ is a Lagrange multiplier satisfying $\mathcal{F}%
^{\nu }\mathcal{=}\ast \mathcal{F}^{\alpha }b_{~\alpha }^{\nu }$ and $M^{2}$
is a two dimensional manifold. In tensor notation (14) becomes

\begin{equation}
S=\frac{1}{2}\int d^{2}\xi \varepsilon _{ij}\mathcal{F}^{i\mu }\mathcal{U}%
_{\mu }^{j}-\frac{1}{4}\int d^{2}\xi \varepsilon _{ij}\mathcal{F}^{i\mu }%
\mathcal{F}^{j\nu }b_{\mu \nu }.  \tag{15}
\end{equation}%
Solving (15) for $\mathcal{F}_{i}^{\mu }$ leads to

\begin{equation}
\mathcal{F}^{i\mu }=-\mathcal{U}_{\nu }^{i}b^{\mu \nu },  \tag{16}
\end{equation}%
where $b^{\mu \nu }=g^{\mu \alpha }g^{\nu \beta }b_{\alpha \beta }$. Thus,
substituting (16) back into (15) yields

\begin{equation}
S_{x}=\frac{1}{4}\int d^{2}\xi \varepsilon _{ij}\mathcal{U}_{\mu }^{i}%
\mathcal{U}_{\nu }^{j}b^{\mu \nu },  \tag{17}
\end{equation}%
or

\begin{equation}
S_{x}=-\frac{1}{4}\int_{M^{2}}\mathcal{U}_{\mu }\wedge \ast \mathcal{U}_{\nu
}g^{\mu \nu },  \tag{18}
\end{equation}%
in abstract notation. Here, in order to emphasize the $x$ depends of $%
\mathcal{U}_{\mu }^{i}$ we wrote $S_{x}$ instead of $S$. Of course (17)
makes sense if one assumes $\mathcal{U}_{\mu }^{i}\neq 0$.

Substituting (12) into (17) one gets%
\begin{equation}
S_{x}=\frac{1}{4}\int d^{2}\xi \varepsilon _{ij}(\mathcal{G}_{\mu \alpha
}^{ik}\frac{\partial x^{\alpha }}{\partial \xi ^{k}})(\mathcal{G}_{\nu \beta
}^{jl}\frac{\partial x^{\beta }}{\partial \xi ^{l}})b^{\mu \nu }.  \tag{19}
\end{equation}%
Since we have the identity

\begin{equation}
\varepsilon _{ij}\mathcal{G}_{\mu \alpha }^{ik}\mathcal{G}_{\nu \beta
}^{jl}b^{\mu \nu }=2\mathcal{G}_{\alpha \beta }^{kl},  \tag{20}
\end{equation}%
we discover that (19) becomes

\begin{equation}
S_{x}=\frac{1}{2}\int d^{2}\xi \mathcal{G}_{\alpha \beta }^{kl}\frac{%
\partial x^{\alpha }}{\partial \xi ^{k}}\frac{\partial x^{\beta }}{\partial
\xi ^{l}},  \tag{21}
\end{equation}%
which using (13) can also be rewritten as%
\begin{equation}
S_{x}=\frac{1}{2}\int d^{2}\xi \delta ^{ij}\frac{\partial x^{\mu }}{\partial
\xi ^{i}}\frac{\partial x^{\nu }}{\partial \xi ^{j}}g_{\mu \nu }+\frac{1}{2}%
\int d^{2}\xi \varepsilon ^{ij}\frac{\partial x^{\mu }}{\partial \xi ^{i}}%
\frac{\partial x^{\nu }}{\partial \xi ^{j}}b_{\mu \nu }.  \tag{22}
\end{equation}%
We recognize in (22) the two dimensional $\sigma $-model action.

The above method can be generalized simply by changing the flat metric $%
\delta _{ij}$ to a world-sheet metric $\gamma _{ij}$. In fact, in this case
(11) (or (12)) can be written as

\begin{equation}
\mathcal{U}_{\mu }^{i}=\gamma ^{ij}\frac{\partial x^{\nu }}{\partial \xi ^{j}%
}g_{\mu \nu }+\epsilon ^{ij}\frac{\partial x^{\nu }}{\partial \xi ^{j}}%
b_{\mu \nu },  \tag{23}
\end{equation}%
where $\epsilon ^{ij}=\frac{\varepsilon ^{ij}}{\sqrt{\gamma }}$. Thus,
following the same procedure we find that the generalized form of (22) is

\begin{equation}
S_{x}=\frac{1}{2}\int d^{2}\xi \sqrt{\gamma }\gamma ^{ij}\frac{\partial
x^{\mu }}{\partial \xi ^{i}}\frac{\partial x^{\nu }}{\partial \xi ^{j}}%
g_{\mu \nu }+\frac{1}{2}\int d^{2}\xi \sqrt{\gamma }\epsilon ^{ij}\frac{%
\partial x^{\mu }}{\partial \xi ^{i}}\frac{\partial x^{\nu }}{\partial \xi
^{j}}b_{\mu \nu }.  \tag{24}
\end{equation}%
In order to recall that (24) was obtained by using the symplectic relation
(2) we shall call the two dimensional action (24) `symplectic-$\sigma $%
-model action'.

Now, from Duff's work [4] we already know how to dualize $S_{x}$. In fact,
the dual action of (24) is

\begin{equation}
S_{y}=\frac{1}{2}\int d^{2}\xi \sqrt{\gamma }\gamma ^{ij}\frac{\partial
y_{\mu }}{\partial \xi ^{i}}\frac{\partial y_{\nu }}{\partial \xi ^{j}}%
p^{\mu \nu }+\frac{1}{2}\int d^{2}\xi \sqrt{\gamma }\epsilon ^{ij}\frac{%
\partial y_{\mu }}{\partial \xi ^{i}}\frac{\partial y_{\nu }}{\partial \xi
^{j}}q^{\mu \nu }.  \tag{25}
\end{equation}%
Here, $p^{\mu \nu }=p^{\nu \mu }$ and $q^{\mu \nu }=-q^{\nu \mu }$ are
related to $g_{\mu \nu }$ and $b_{\mu \nu }$ by the expressions

\begin{equation}
p=(g-bg^{-1}b)^{-1}  \tag{26}
\end{equation}%
and

\begin{equation}
q=-g^{-1}b(g-bg^{-1}b)^{-1}.  \tag{27}
\end{equation}%
One can show that the field equations of the action $S_{x}$ are the Bianchi
identities for the dual action $S_{y}$ and the Bianchi identities of the
original action $S_{x}$ are the field equations for $S_{y}$. In fact, one
finds that the coordinates $x^{\mu }$ and $y_{\mu }$ are related by

\begin{equation}
\epsilon ^{ij}\frac{\partial y_{\mu }}{\partial \xi ^{j}}=\gamma ^{ij}\frac{%
\partial x^{\nu }}{\partial \xi ^{j}}g_{\mu \nu }+\epsilon ^{ij}\frac{%
\partial x^{\nu }}{\partial \xi ^{j}}b_{\mu \nu }  \tag{28}
\end{equation}%
and

\begin{equation}
\epsilon ^{ij}\frac{\partial x^{\nu }}{\partial \xi ^{j}}=\gamma ^{ij}\frac{%
\partial y_{\nu }}{\partial \xi ^{j}}p^{\mu \nu }+\epsilon ^{ij}\frac{%
\partial y_{\nu }}{\partial \xi ^{j}}q^{\mu \nu }.  \tag{29}
\end{equation}%
Moreover, (28) and (29) can be united into a single equation

\begin{equation}
\Omega _{MN}\gamma ^{ij}\frac{\partial Z^{N}}{\partial \xi ^{j}}%
=G_{MN}\epsilon ^{ij}\frac{\partial Z^{N}}{\partial \xi ^{j}},  \tag{30}
\end{equation}%
where $Z^{M}=(x^{\mu },y_{\alpha })$, with $M=1,...,2n$ and $\Omega $ and $G$
are given by the symmetric matrices

\begin{equation}
\Omega _{MN}=\left( 
\begin{array}{cc}
0 & \delta _{\mu }^{\alpha } \\ 
\delta _{\alpha }^{\mu } & 0%
\end{array}%
\right) ,  \tag{31}
\end{equation}%
and

\begin{equation}
G_{MN}=\left( 
\begin{array}{cc}
g_{\mu \nu }-b_{\mu \alpha }g^{\alpha \beta }b_{\nu \beta } & b_{\mu
}^{~~\alpha } \\ 
-b_{~~\nu }^{\beta } & g^{\alpha \beta }%
\end{array}%
\right) .  \tag{32}
\end{equation}%
Expression (30) shows explicitly the dual $SO(n,n)$-symmetry (see Ref. [4]
for details).

In order to find a possible connection between this Duff's dual formalism
and the above discussion on symplectic-$\sigma $-model action we shall
define the object

\begin{equation}
\mathcal{P}_{ij}^{\mu \nu }=\gamma _{ij}p^{\mu \nu }+\epsilon _{ij}q^{\mu
\nu }.  \tag{33}
\end{equation}%
Let us assume the expression

\begin{equation}
\mathcal{P}_{jk}^{\mu \alpha }\mathcal{G}_{\alpha \nu }^{ki}=\delta
_{j}^{i}\delta _{\nu }^{\mu }.  \tag{34}
\end{equation}%
Explicitly, one has

\begin{equation}
(\gamma _{jk}p^{\mu \alpha }+\epsilon _{jk}q^{\mu \alpha })(\gamma
^{ki}g_{\alpha \nu }+\epsilon ^{ki}b_{\alpha \nu })=\delta _{j}^{i}\delta
_{\nu }^{\mu }.  \tag{35}
\end{equation}%
From this formula one can derive the two equations

\begin{equation}
p_{~~\nu }^{\mu }-q^{\mu \alpha }b_{\alpha \nu }=\delta _{\nu }^{\mu } 
\tag{36}
\end{equation}%
and

\begin{equation}
q_{~~\nu }^{\mu }+p^{\mu \alpha }b_{\alpha \nu }=0.  \tag{37}
\end{equation}%
Substituting (37) into (36) yields%
\begin{equation}
p^{\mu \alpha }(g_{\alpha \nu }-b_{\alpha \tau }g^{\tau \lambda }b_{\nu
\lambda })=\delta _{\nu }^{\mu },  \tag{38}
\end{equation}%
which leads to the symbolic expression (26). Similarly, by substituting (36)
into (37) one finds

\begin{equation}
q^{\mu \alpha }(g_{\alpha \nu }-b_{\alpha \tau }g^{\tau \lambda }b_{\nu
\lambda })=-g^{\mu \alpha }b_{\alpha \nu }.  \tag{39}
\end{equation}%
When written in a symbolic form this expression leads to (27). Moreover, let
us denote by $(b^{-1})^{\mu \nu }$ the inverse of $b_{\mu \nu }$, that is,
we have $(b^{-1})^{\mu \alpha }b_{\alpha \nu }=\delta _{\nu }^{\mu }$.
Therefore, multiplying (39) by $b^{-1}g$ one obtains

\begin{equation}
q^{\mu \alpha }(b_{\alpha \rho }+g_{\alpha \nu }(b^{-1})^{\nu \tau }g_{\tau
\rho })=-\delta _{\rho }^{\mu }.  \tag{40}
\end{equation}%
It is important to emphasize that in general $(b^{-1})^{\mu \nu }\neq b^{\mu
\nu }=g^{\mu \alpha }g^{\nu \beta }b_{\alpha \beta }$.

Now it is straightforward to see from (38) and (39) that the Duff formalism
makes sense only when

\begin{equation}
g_{\alpha \nu }-b_{\alpha \tau }g^{\tau \lambda }b_{\nu \lambda }\neq 0. 
\tag{41}
\end{equation}%
Of course, the same conclusion could be obtained observing that the symbolic
equations (26) and (27) are consistent only if one assumes the relation $%
g-bg^{-1}b\neq 0$ (which is the symbolic form of (41)). Nevertheless, the
explicit formula (41) may help to clarify this observation.

In the case of a symplectic-$\sigma $-model we have assumed the formula $%
b_{~\beta }^{\mu }b_{~\nu }^{\beta }=-\delta _{\nu }^{\mu }$, given in (2).
This relation is equivalent to the expression

\begin{equation}
g_{\alpha \nu }-b_{\alpha \tau }g^{\tau \lambda }b_{\nu \lambda }=0. 
\tag{42}
\end{equation}%
Comparing (41) and (42) one discovers that one can not use the Duff's
prescription in the case of a symplectic-$\sigma $-model. Consequently, the
elegant formula (30) does not follow from a symplectic-$\sigma $-model and
therefore we can not describe the dual symplectic-$\sigma $-model theory in
terms of a manifest $SO(n,n)$-symmetry. This is even more evident if we
observe that the "Kaluza-Klein" type metric (32) becomes singular when the
formula (42) is assumed.

It is interesting to observe that by assuming (2) or (42) one has $%
(b^{-1})^{\mu \nu }=-b^{\mu \nu }$ and therefore the the expression (40) is
not consistent.

The above analysis means that it does not exist a dual theory for the case
of symplectic-$\sigma $-model at least with manifest $SO(n,n)$-symmetry. A
possible explanation for this result may be described as follows. Consider
the a subspace $A$ of the Hilbert space of holomorphic differentials $H$.
One can show that $H$=$A\oplus \bar{A}$, where $\bar{A}$ is the space
complex conjugate differentials. In fact, it turns out that the complex
conjugate operator $-$ defines an isomorphism of $A$ onto $\bar{A}$, so dim $%
A$=dim$\bar{A}$. It seems to us that this isomorphism defines some kind of
self-duality for the complex structure. So, from this point of view there is
not dual symplectic-$\sigma $-model because the model itself is self-dual.

It is interesting to mention that an alternative approach for a possible
dual theory can be obtained by writing the first order action [5]-[7] (see
also Ref. [8-10])

\begin{equation}
\begin{array}{c}
S_{(x,y)}=\frac{1}{2}\int d^{2}\xi \sqrt{\gamma }\gamma ^{ij}(\frac{\partial
x^{\mu }}{\partial \xi ^{i}}-A_{i}^{\mu })(\frac{\partial x^{\nu }}{\partial
\xi ^{j}}-A_{j}^{\nu })g_{\mu \nu } \\ 
\\ 
+\frac{1}{2}\int d^{2}\xi \sqrt{\gamma }\epsilon ^{ij}(\frac{\partial x^{\mu
}}{\partial \xi ^{i}}-A_{i}^{\mu })(\frac{\partial x^{\nu }}{\partial \xi
^{j}}-A_{j}^{\nu })b_{\mu \nu }+\int d^{2}\xi \sqrt{\gamma }\epsilon ^{ij}%
\frac{\partial y_{\mu }}{\partial \xi ^{i}}A_{j}^{\mu }.%
\end{array}
\tag{43}
\end{equation}%
From this action one may attempt to obtain the two actions (24) and (25). In
principle, this program can be achieved following two separated steps. First
varying (43) with respect to $y_{\mu }$ one observe that (43) implies $dA=0$%
. This means that we can set $A=0$ and therefore the action (43) is reduced
to (24). On the other hand by setting $x^{\mu }=0$ in (43) we get the the
reduced first order action

\begin{equation}
\begin{array}{c}
S_{(0,y)}=\frac{1}{2}\int d^{2}\xi \sqrt{\gamma }\gamma ^{ij}A_{i}^{\mu
}A_{j}^{\nu }g_{\mu \nu } \\ 
\\ 
+\frac{1}{2}\int d^{2}\xi \sqrt{\gamma }\epsilon ^{ij}A_{i}^{\mu }A_{j}^{\nu
}b_{\mu \nu }+\int d^{2}\xi \sqrt{\gamma }\epsilon ^{ij}\frac{\partial
y_{\mu }}{\partial \xi ^{i}}A_{j}^{\mu }.%
\end{array}
\tag{44}
\end{equation}%
Varying with respect $A_{i}^{\mu }$ we obtain the equation

\begin{equation}
\epsilon ^{ij}\frac{\partial y_{\mu }}{\partial \xi ^{j}}=\gamma
^{ij}A_{i}^{\mu }g_{\mu \nu }+\epsilon ^{ij}A_{i}^{\mu }b_{\mu \nu }. 
\tag{45}
\end{equation}%
The idea then will be to solve (45) for $A_{i}^{\mu }$ in terms of $\frac{%
\partial y_{\mu }}{\partial \xi ^{j}}$ and to substitute the result back
into (44). In principle with this method one should be able to obtain the
action (25). However, in the case of the symplectic-$\sigma $-model this is
not possible as we have shown. So, the action (43) does not solve the
problem of finding the dual symplectic-$\sigma $-model. Nevertheless,
something interesting may arise by considering (43) from another point of
view as we now explain.

In the usual case under compactification the action (43) may transfer
space-like coordinates in the $x$ scenario to space-like coordinates $y$ in
dual theory and \textit{vice versa}. But this phenomena depends whether we
choose the Euclidean metric

\begin{equation}
\delta _{ij}=\left( 
\begin{array}{cc}
1 & 0 \\ 
0 & 1%
\end{array}%
\right) ,  \tag{46}
\end{equation}%
or the Minkowski metric 
\begin{equation}
\eta _{ij}=\left( 
\begin{array}{cc}
1 & 0 \\ 
0 & -1%
\end{array}%
\right) .  \tag{47}
\end{equation}%
(see Ref. [5] for details). This is important because at the level of a
possible action if we choose $\delta _{ij}$ we have $\sigma $-model action,
while if we choose $\eta _{ij}$ we get the string theory action.
Traditionally, these results are not really a problem because one can take
recourse of a Wick rotation. However, at a more fundamental level one takes
either $\delta _{ij}$ or $\eta _{ij}$ but not both. The situation is even
more intriguing in the case of a symplectic-$\sigma $-model because in such
a case one uses $\delta _{ij}$ instead of $\eta _{ij}$. These observations
motive us to review the complex numbers structure.

As it is known a complex number can be written as

\begin{equation}
\left( 
\begin{array}{cc}
x & y \\ 
-y & x%
\end{array}%
\right) =x\left( 
\begin{array}{cc}
1 & 0 \\ 
0 & 1%
\end{array}%
\right) +y\left( 
\begin{array}{cc}
0 & 1 \\ 
-1 & 0%
\end{array}%
\right) .  \tag{48}
\end{equation}%
Since $\left( 
\begin{array}{cc}
0 & 1 \\ 
-1 & 0%
\end{array}%
\right) \left( 
\begin{array}{cc}
0 & 1 \\ 
-1 & 0%
\end{array}%
\right) =-\left( 
\begin{array}{cc}
1 & 0 \\ 
0 & 1%
\end{array}%
\right) $ one finds that the matrix $\left( 
\begin{array}{cc}
0 & 1 \\ 
-1 & 0%
\end{array}%
\right) $ can be identified with the usual imaginary unit $i$.

It turns out that the matrices $\left( 
\begin{array}{cc}
1 & 0 \\ 
0 & 1%
\end{array}%
\right) $ and $\left( 
\begin{array}{cc}
0 & 1 \\ 
-1 & 0%
\end{array}%
\right) $ can be considered as two of the matrix bases of general real $%
2\times 2$ matrices which we denote by $M(2,R)$. In fact, any $2\times 2$
matrix $\Gamma $ over the real can be written as

\begin{equation}
\begin{array}{c}
\Gamma =\left( 
\begin{array}{cc}
a & b \\ 
c & d%
\end{array}%
\right) =x\left( 
\begin{array}{cc}
1 & 0 \\ 
0 & 1%
\end{array}%
\right) +y\left( 
\begin{array}{cc}
0 & 1 \\ 
-1 & 0%
\end{array}%
\right) \\ 
\\ 
+r\left( 
\begin{array}{cc}
1 & 0 \\ 
0 & -1%
\end{array}%
\right) +s\left( 
\begin{array}{cc}
0 & 1 \\ 
1 & 0%
\end{array}%
\right) ,%
\end{array}
\tag{49}
\end{equation}%
where

\begin{equation}
\begin{array}{ccc}
x=\frac{1}{2}(a+d), &  & y=\frac{1}{2}(b-c), \\ 
&  &  \\ 
r=\frac{1}{2}(a-d), &  & s=\frac{1}{2}(b+c).%
\end{array}
\tag{50}
\end{equation}

Let us rewrite (49) in the form

\begin{equation}
\Gamma _{ij}=x\delta _{ij}+y\varepsilon _{ij}+r\eta _{ij}+s\lambda _{ij}, 
\tag{51}
\end{equation}%
where

\begin{equation}
\begin{array}{ccc}
\delta _{ij}\equiv \left( 
\begin{array}{cc}
1 & 0 \\ 
0 & 1%
\end{array}%
\right) , &  & \varepsilon _{ij}\equiv \left( 
\begin{array}{cc}
0 & 1 \\ 
-1 & 0%
\end{array}%
\right) , \\ 
&  &  \\ 
\eta _{ij}\equiv \left( 
\begin{array}{cc}
1 & 0 \\ 
0 & -1%
\end{array}%
\right) , &  & \lambda _{ij}\equiv \left( 
\begin{array}{cc}
0 & 1 \\ 
1 & 0%
\end{array}%
\right) .%
\end{array}
\tag{52}
\end{equation}%
Considering this notation, we find that (48) becomes

\begin{equation}
z_{ij}=x\delta _{ij}+y\varepsilon _{ij}.  \tag{53}
\end{equation}%
Observe that (53) can be obtained from (51) by setting $r=0$ and $s=0.$ If $%
ad-bc\neq 0,$ that is if $\det \Gamma \neq 0$, then the matrices in $M(2,R)$
can be associated with the group $GL(2,R)$. If in addition we require $%
ad-bc=1$, then one gets the elements of the subgroup $SL(2,R)$.

Traditionally, one may start a gravitational theory by choosing a flat
metric with Euclidean signature or Minkowski signature, but not both.
However, from the point of view of $2\times 2$ matrices both signatures are
equally important. So, if we choose Euclidean signature $\delta _{ij}$ and
the matrix $\varepsilon _{ij}$ we may have the complex structure (53). Then
the question arises why not to choose instead the Minkowski metric $\eta
_{ij}$ and the matrix $\lambda _{ij}$? In such a case one should have the
alternative numbers

\begin{equation}
\omega _{ij}=r\eta _{ij}+s\lambda _{ij}.  \tag{54}
\end{equation}%
The immediate answer to the above question can be expressed saying that the
algebra corresponding to the possible number (54) is not closed. In fact, as
a matrices $\eta _{ij}$ and $\lambda _{ij}$ satisfy the algebra

\begin{equation}
\left( 
\begin{array}{cc}
1 & 0 \\ 
0 & -1%
\end{array}%
\right) \left( 
\begin{array}{cc}
1 & 0 \\ 
0 & -1%
\end{array}%
\right) =\left( 
\begin{array}{cc}
1 & 0 \\ 
0 & 1%
\end{array}%
\right) ,  \tag{55}
\end{equation}%
\begin{equation}
\left( 
\begin{array}{cc}
1 & 0 \\ 
0 & -1%
\end{array}%
\right) \left( 
\begin{array}{cc}
0 & 1 \\ 
1 & 0%
\end{array}%
\right) =\left( 
\begin{array}{cc}
0 & 1 \\ 
1 & 0%
\end{array}%
\right) ,  \tag{56}
\end{equation}%
\begin{equation}
\left( 
\begin{array}{cc}
0 & 1 \\ 
1 & 0%
\end{array}%
\right) \left( 
\begin{array}{cc}
0 & 1 \\ 
1 & 0%
\end{array}%
\right) =\left( 
\begin{array}{cc}
1 & 0 \\ 
0 & 1%
\end{array}%
\right) .  \tag{57}
\end{equation}%
However something interesting arises if we analyze this algebra from tensor
analysis point of view.

The same algebra (55)-(57) can be written as

\begin{equation}
\eta _{ik}\delta ^{kl}\eta _{lj}=\delta _{ij},  \tag{58}
\end{equation}%
\begin{equation}
\eta _{ik}\delta ^{kl}\lambda _{lj}=\varepsilon _{ij},  \tag{59}
\end{equation}%
\begin{equation}
\lambda _{ik}\delta ^{kl}\lambda _{lj}=\delta _{ij}.  \tag{60}
\end{equation}%
Now, from (58)-(60) it is evident that we are combining the two flat metrics 
$\eta _{ik}$ and $\delta _{ij}$ instead to consider either $\eta _{ik}$ or $%
\delta _{ij}$. This may be interesting for some kind of two dimensional
gravitational theory but in principle what we would like is to choose either 
$\eta _{ik}$ or $\delta _{ij}$, but not both. So, an alternative algebra
will be

\begin{equation}
\eta _{ik}\eta ^{kl}\eta _{lj}=\eta _{ij},  \tag{61}
\end{equation}%
\begin{equation}
\eta _{ik}\eta ^{kl}\lambda _{lj}=\lambda _{ij},  \tag{62}
\end{equation}%
\begin{equation}
\lambda _{ik}\eta ^{kl}\lambda _{lj}=-\eta _{ij},  \tag{63}
\end{equation}%
which is closed. So, with this algebra we can perfectly use the numbers in
(54). Just to recall that this numbers can be related to a matroid theory we
shall call such numbers complexoids. The reason for this comes from the
observation in the Ref. [11] that the matrices in (52) can be linked to a
2-rank self-dual representable matroid $\mathcal{M}=(E,\mathcal{B})$ via the
matrix

\begin{equation}
V_{i}^{A}=\left( 
\begin{array}{cccc}
1 & 0 & 0 & 1 \\ 
0 & 1 & -1 & 0%
\end{array}%
\right) ,  \tag{64}
\end{equation}%
with the index $A$ taking values in the set 
\begin{equation}
E=\{1,2,3,4\}.  \tag{65}
\end{equation}%
It turns out that the subsets $\{ \mathbf{V}^{1},\mathbf{V}^{2}\}$, $\{%
\mathbf{V}^{1},\mathbf{V}^{3}\}$, $\{ \mathbf{V}^{2},\mathbf{V}^{4}\}$ and $%
\{ \mathbf{V}^{3},\mathbf{V}^{4}\}$ are bases over the real of the matrix
(64). One can associate with these subsets the collection%
\begin{equation}
\mathcal{B}=\{ \{1,2\},\{1,3\},\{2,4\},\{3,4\} \},  \tag{66}
\end{equation}%
which is a family of subsets of $E$. It can be shown that $\mathcal{M}=(E,%
\mathcal{B})$ is graphic and orientable. In the later case the corresponding
chirotope is given by

\begin{equation}
\chi ^{AB}=\varepsilon ^{ij}V_{i}^{A}V_{j}^{B}.  \tag{67}
\end{equation}%
Thus, we get, as nonvanishing elements of the chirotope $\chi ^{AB}$, the
combinations

\begin{equation}
\begin{array}{cccc}
12+, & 13-, & 24-, & 34+.%
\end{array}
\tag{68}
\end{equation}%
The relation of this matroid structure with (52) comes from the
identification $\{ \mathbf{V}^{1},\mathbf{V}^{2}\} \rightarrow \delta _{ij}$%
, $\{ \mathbf{V}^{1},\mathbf{V}^{3}\} \rightarrow \eta _{ij}$, $\{ \mathbf{V}%
^{2},\mathbf{V}^{4}\} \rightarrow \lambda _{ij}$ and $\{ \mathbf{V}^{3},%
\mathbf{V}^{4}\} \rightarrow \varepsilon _{ij}$. The signs in (68)
correspond to the determinants of the matrices $\delta _{ij}$, $\eta _{ij}$, 
$\lambda _{ij}$ and $\varepsilon _{ij}$, which can be calculated using (67)
(see Ref. [11] for details (see also Refs. [12]-[14] and references
therein)).

In the case of complexoids we shall have the analogue of Cauchy-Riemann
conditions

\begin{equation}
\mathcal{V}_{\mu }^{i}=\eta ^{ij}\frac{\partial x^{\nu }}{\partial \xi ^{j}}%
g_{\mu \nu }+\lambda ^{ij}\frac{\partial x^{\nu }}{\partial \xi ^{j}}b_{\mu
\nu }=0,  \tag{69}
\end{equation}%
where we change $\delta ^{ij}$ by $\eta ^{ij}$ and $\lambda ^{ij}$ by $%
\varepsilon ^{ij}$ in (11). In this case one may propose the analogue of the
action (17), namely

\begin{equation}
\mathcal{S}_{x}=\frac{1}{4}\int d\xi ^{2}\lambda _{ij}\mathcal{V}_{\mu }^{i}%
\mathcal{V}_{\nu }^{j}b^{\mu \nu },  \tag{70}
\end{equation}%
with $\mathcal{V}_{\mu }^{i}\neq 0$. In this case, of course, one must
assume $b_{\mu \nu }=b_{\nu \mu }$. In fact, just as (17) leads to (22) one
can show that by considering the algebra (61)-(63) that (70) leads to the
action

\begin{equation}
\mathcal{S}_{x}=\frac{1}{2}\int d\xi ^{2}\eta ^{ij}\frac{\partial x^{\mu }}{%
\partial \xi ^{i}}\frac{\partial x^{\nu }}{\partial \xi ^{j}}g_{\mu \nu }+%
\frac{1}{2}\int d\xi ^{2}\lambda ^{ij}\frac{\partial x^{\mu }}{\partial \xi
^{i}}\frac{\partial x^{\nu }}{\partial \xi ^{j}}b_{\mu \nu }.  \tag{71}
\end{equation}%
This is of course a type of string theory action with a (1+1)-world-sheet
flat metric $\eta ^{ij}$, instead of (0+2)-world-sheet Euclidean flat metric 
$\delta ^{ij}$, as it is the case when one is using a complex structure.
Thus, without the need of using Wick rotation we were able to obtain (71).
Indeed, the interesting thing is that just as (22) can be related to complex
structure the action (71) can be linked to the new type of numbers which we
called complexoids.

A natural generalization of (69) and (71) can be obtained by the
transformation $\eta ^{ij}\rightarrow \gamma ^{ij}$. In this way from (71)
we get exactly the same action (24). However, one needs to keep in mind that
in this case the world-sheet metric $\gamma _{ij}$ is associated with the
Minkowski metric $\eta _{ij}$ rather than with the Euclidean metric $\delta
_{ij}$.

We still need to justify formula (69). For this purpose, one may first
recall the traditional method, in the usual complex numbers theory, for
obtaining the Cauchy-Riemann equations. In such case, one defines the
derivative $f^{\prime }$ of a complex valued function $f(\xi ^{i})=x^{1}(\xi
^{i})+ix^{2}(\xi ^{i})$ as an ifinitesimal limit $\bigtriangleup \xi
^{i}\rightarrow 0$ of the ratio

\begin{equation}
\frac{\bigtriangleup f(\xi ^{i})}{\bigtriangleup \xi ^{i}}=\frac{%
\bigtriangleup x^{1}+i\bigtriangleup x^{2}}{\bigtriangleup \xi
^{1}+i\bigtriangleup \xi ^{2}}.  \tag{72}
\end{equation}%
Assuming $\bigtriangleup \xi ^{2}=0$ one gets

\begin{equation}
\frac{\bigtriangleup f(\xi ^{i})}{\bigtriangleup \xi ^{i}}=\frac{%
\bigtriangleup x^{1}+i\bigtriangleup x^{2}}{\bigtriangleup \xi ^{1}}, 
\tag{73}
\end{equation}%
that is, we have

\begin{equation}
f^{\prime }=\frac{\partial x^{1}}{\partial \xi ^{1}}+i\frac{\partial x^{2}}{%
\partial \xi ^{1}}.  \tag{74}
\end{equation}%
But if one assumes $\bigtriangleup \xi ^{1}=0$ one finds

\begin{equation}
\frac{\bigtriangleup f(\xi ^{i})}{\bigtriangleup \xi ^{i}}=\frac{%
\bigtriangleup x^{1}+i\bigtriangleup x^{2}}{i\bigtriangleup \xi ^{2}}. 
\tag{75}
\end{equation}%
This gives

\begin{equation}
f^{\prime }=-i\frac{\partial x^{1}}{\partial \xi ^{2}}+\frac{\partial x^{2}}{%
\partial \xi ^{2}}.  \tag{76}
\end{equation}%
Of course the derivative $f^{\prime }$ makes sense if the expressions (74)
and (76) are the same. This leads to the Cauchy-Riemann equations

\begin{equation}
\begin{array}{c}
\frac{\partial x^{1}}{\partial \xi ^{1}}=\frac{\partial x^{2}}{\partial \xi
^{2}}, \\ 
\\ 
\frac{\partial x^{1}}{\partial \xi ^{2}}=-\frac{\partial x^{2}}{\partial \xi
^{1}}.%
\end{array}
\tag{77}
\end{equation}%
In the matrices notation, if one consider the plane $R^{2}$ and the matrices 
$\delta $ and $\varepsilon $ given in (52), that is if one assumes the
triplite $(R^{2},\delta ,\varepsilon )$, one may write (72) in the symbolic
form

\begin{equation}
\frac{\bigtriangleup f(\xi ^{i})}{\bigtriangleup \xi ^{i}}=\frac{%
\bigtriangleup x^{1}\delta +\bigtriangleup x^{2}\varepsilon }{\bigtriangleup
\xi ^{1}\delta +\bigtriangleup \xi ^{2}\varepsilon },  \tag{78}
\end{equation}%
So, one sees that (73) and (75) become

\begin{equation}
\frac{\bigtriangleup f(\xi ^{i})}{\bigtriangleup \xi ^{i}}=\frac{%
\bigtriangleup x^{1}\delta +\bigtriangleup x^{2}\varepsilon }{\bigtriangleup
\xi ^{1}\delta }  \tag{79}
\end{equation}%
and%
\begin{equation}
\frac{\bigtriangleup f(\xi ^{i})}{\bigtriangleup \xi ^{i}}=\frac{%
\bigtriangleup x^{1}\delta +\bigtriangleup x^{2}\varepsilon }{\bigtriangleup
\xi ^{2}\varepsilon },  \tag{80}
\end{equation}%
respectively. From (79) we find

\begin{equation}
f^{\prime }=\frac{1}{\delta }(\frac{\partial x^{1}}{\partial \xi ^{1}}\delta
+\frac{\partial x^{2}}{\partial \xi ^{1}}\varepsilon ).  \tag{81}
\end{equation}%
On the other hand, (76) is obtained from (79) when one computes

\begin{equation}
\frac{\bigtriangleup f(\xi ^{i})}{\bigtriangleup \xi ^{i}}=\frac{\varepsilon
\delta (\bigtriangleup x^{1}\delta +\bigtriangleup x^{2}\varepsilon )}{%
\varepsilon \delta (\bigtriangleup \xi ^{2}\varepsilon )}.  \tag{82}
\end{equation}%
In fact, one finds

\begin{equation}
f^{\prime }=\frac{1}{\delta }(-\varepsilon \frac{\partial x^{1}}{\partial
\xi ^{2}}+\delta \frac{\partial x^{2}}{\partial \xi ^{2}}).  \tag{83}
\end{equation}%
Once again (81) and (83) lead the Cauchy-Riemann conditions (77).

Now, consider the complexoid structure ($R^{2},\eta ,\lambda $) where $\eta $
and $\lambda $ are given in (52). In this case, one must have

\begin{equation}
\frac{\bigtriangleup f(\xi ^{i})}{\bigtriangleup \xi ^{i}}=\frac{%
\bigtriangleup x^{1}\eta +\bigtriangleup x^{2}\lambda }{\bigtriangleup \xi
^{1}\eta +\bigtriangleup \xi ^{2}\lambda }.  \tag{84}
\end{equation}%
When $\bigtriangleup \xi ^{2}=0$ one gets

\begin{equation}
\frac{\bigtriangleup f(\xi ^{i})}{\bigtriangleup \xi ^{i}}=\frac{%
\bigtriangleup x^{1}\eta +\bigtriangleup x^{2}\lambda }{\bigtriangleup \xi
^{1}\eta },  \tag{85}
\end{equation}%
while when $\bigtriangleup \xi ^{1}=0$ one obtains 
\begin{equation}
\frac{\bigtriangleup f(\xi ^{i})}{\bigtriangleup \xi ^{i}}=\frac{%
\bigtriangleup x^{1}\eta +\bigtriangleup x^{2}\lambda }{\bigtriangleup \xi
^{2}\lambda }.  \tag{86}
\end{equation}%
From (85) we find

\begin{equation}
f^{\prime }=\frac{1}{\eta }(\frac{\partial x^{1}}{\partial \xi ^{1}}\eta +%
\frac{\partial x^{2}}{\partial \xi ^{1}}\lambda ).  \tag{87}
\end{equation}%
While writing (86) in the form%
\begin{equation}
\frac{\bigtriangleup f(\xi ^{i})}{\bigtriangleup \xi ^{i}}=\frac{\lambda
\eta (\bigtriangleup x^{1}\eta +\bigtriangleup x^{2}\lambda )}{\lambda \eta
(\bigtriangleup \xi ^{2}\lambda )}  \tag{88}
\end{equation}%
and using the algebra (61)-(63) one obtains the formula

\begin{equation}
f^{\prime }=\frac{1}{\eta }(-\lambda \frac{\partial x^{1}}{\partial \xi ^{2}}%
+\eta \frac{\partial x^{2}}{\partial \xi ^{2}}).  \tag{89}
\end{equation}%
Surprisingly, from (87) and (89) one again obtains the Cauchy-Riemann
equations (77). However one needs to keep in mind that in this case (77)
refers to the complexoid structure ($R^{2},\eta ,\lambda $) rather than the
complex structure ($R^{2},\delta ,\varepsilon $). In fact, this is verified
by writing the Cauchy-Riemann equations in the two different ways

\begin{equation}
\delta ^{ij}\frac{\partial x^{\nu }}{\partial \xi ^{j}}\delta _{\mu \nu
}-\varepsilon ^{ij}\frac{\partial x^{\nu }}{\partial \xi ^{j}}\varepsilon
_{\mu \nu }=0,  \tag{90}
\end{equation}%
and

\begin{equation}
\eta ^{ij}\frac{\partial x^{\nu }}{\partial \xi ^{j}}\eta _{\mu \nu
}-\lambda ^{ij}\frac{\partial x^{\nu }}{\partial \xi ^{j}}\lambda _{\mu \nu
}=0.  \tag{91}
\end{equation}%
Here, the indices $\mu ,\nu $ take values in the set $\{1,2\}$. The
expression (90) can be obtained from (11) when we set $g_{\mu \nu }=\delta
_{\mu \nu }$ and $b_{\mu \nu }=-\varepsilon _{\mu \nu }$, while (91) is
obtained from (69) by setting $g_{\mu \nu }=\eta _{\mu \nu }$ and $b_{\mu
\nu }=-\lambda _{\mu \nu }$. A generalization of (91) is precisely (69).

Further motivation for our approach may arise from the following
observations. It is known that the fundamental matrices $\delta
_{ij},\varepsilon _{ij},\eta _{ij}$ and $\lambda _{ij}$ given in (52) not
only form a basis for $M(2,R)$ but also determine a basis for the Clifford
algebras $C(2,0)$ and $C(1,1)$. In fact one has the isomorphisms $M(2,R)\sim
C(2,0)\sim C(1,1)$. Moreover, one can show that $C(0,2)$ can be constructed
using the fundamental matrices (52) and Kronecker products. In this way all
the others $C(a,b)$'s can be constructed from the building blocks $C(2,0),$ $%
C(1,1)$ and $C(0,2)$. Therefore, in this context by combining complex and
complexoid structures one should expect that interesting relations may
emerge between the Clifford structure and the symplectic-$\sigma $-model
theory.

\bigskip \ 

\begin{center}
\textbf{Acknowledgments}
\end{center}

I would like to thank to M. C. Mar\'{\i}n for helpful comments. This work
was partially supported by PROFAPI 2009 and PIFI 3.3.

\end{document}